\documentclass[12pt]{article}

\usepackage{times}
\usepackage{comment}
\usepackage{xcolor}
\usepackage{empheq}
\usepackage[superscript,nomove]{cite}
\usepackage{amsmath}
\usepackage{amssymb}
\usepackage{ulem,xpatch}
\usepackage{titling}
\usepackage[font={footnotesize}]{caption}
\usepackage{lineno}
\usepackage[utf8]{inputenc}

\newcommand{\bnth}{\ensuremath{\bar n_\mathrm{th}}}
\newcommand{\bncav}{\ensuremath{\bar n_\mathrm{cav}}}
\newcommand{\bnmin}{\ensuremath{\bar n_\mathrm{min}}}

\newcommand{\SFFqba}{\bar S_{FF}^{\mathrm{qba}}}
\newcommand{\SFFth}{\bar S_{FF}^{\mathrm{th}}}
\newcommand{\SxxImp}{\bar S_{xx}^{\mathrm{imp}}}

\newcommand{\SxxAdd}{\bar S_{xx}^{\mathrm{add}}}
\newcommand{\SxxSQL}{\bar S_{xx}^{\mathrm{SQL}}}
\newcommand{\SxxInt}{\bar S_{xx}^{\mathrm{int}}}
\newcommand{\SxxMeas}{\bar S_{xx}^{\mathrm{meas}}}

\newcommand{\ximp}{\hat{x}_{\mathrm{imp}}}

\newcommand{\Fth}{\hat{F}_{\mathrm{th}}}
\newcommand{\Fqba}{\hat{F}_{\mathrm{qba}}}
\newcommand{\xzpf}{x_{\mathrm{zpf}}}

\newcommand{\kappaaux}{\kappa_{\mathrm{aux}}}
\newcommand{\deltaaux}{\Delta_{\mathrm{aux}}}

\newcommand{\chic}{\chi_\mathrm{c}}
\newcommand{\chim}{\chi_\mathrm{m}}
\newcommand{\chieff}{\chi_\mathrm{eff}}

\newcommand{\Og}{\Omega}
\newcommand{\Om}{\Omega_\mathrm{m}}
\newcommand{\Oeff}{\Omega_\mathrm{eff}}
\newcommand{\Gm}{\Gamma_\mathrm{m}}
\newcommand{\Gmeas}{\Gamma_{\mathrm{meas}}}
\newcommand{\Gqba}{\Gamma_{\mathrm{qba}}}

\newcommand{\etad}{\eta_\mathrm{det}}

\newcommand{\Cq}{\ensuremath{C_\mathrm{q}}}

\usepackage{abstract}

\title{\bf\vspace{-2cm}\textsf{\Large Continuous Force and Displacement Measurement Below the Standard Quantum Limit}}
\author{\normalsize{David Mason$^{1,2,*}$, Junxin Chen$^{1,2,*}$, Massimiliano Rossi$^{1,2,*}$, Yeghishe Tsaturyan$^{1}$ \& \\Albert Schliesser$^{1,2,\dagger}$}\\
\vspace{7mm}\small{\it $^{1}$Niels Bohr Institute, University of Copenhagen, 2100 Copenhagen, Denmark}\\
\small{\it$^{2}$Center for Hybrid Quantum Networks (Hy-Q), Niels Bohr Institute,}\\
\small{\it University of Copenhagen, 2100 Copenhagen, Denmark}\\
\vspace{7mm}\small{$^\ast$these authors contributed equally to this work}\\
\small{$^\dagger$to whom correspondence should be addressed; e-mail:  albert.schliesser@nbi.ku.dk}}
\date{}

\pretitle{\begin{flushleft}\LARGE} 
\posttitle{\end{flushleft}}
\preauthor{\begin{flushleft}\large} 
\postauthor{\end{flushleft}}

\begin{document}

\maketitle

\begin{abstract}
\textbf{
Quantum mechanics dictates that the precision of physical measurements must be subject to certain constraints. In the case of inteferometric displacement measurements, these restrictions impose a `standard quantum limit' (SQL), which optimally balances the precision of a measurement with its unwanted backaction\cite{Braginskii1968}.  To go beyond this limit, one must devise more sophisticated measurement techniques, which either `evade' the backaction of the measurement\cite{Braginsky1980}, or achieve clever cancellation of the unwanted noise at the detector\cite{Unruh1982,Vyatchanin1995}.  In the half-century since the SQL was established, systems ranging from LIGO\cite{Aasi2013} to ultracold atoms\cite{Schreppler2014} and nanomechanical devices\cite{LaHaye2004,Rossi2018} have pushed displacement measurements towards this limit, and a variety of sub-SQL techniques have been tested in proof-of-principle experiments\cite{Kampel2017,Suh2014,Wollman2015b,Lecocq2015,Sudhir2017a}.  However, to-date, no experimental system has successfully demonstrated an interferometric displacement measurement with sensitivity (including \textit{all} relevant noise sources: thermal, backaction, and imprecision) below the SQL.  Here, we exploit strong quantum correlations in an ultracoherent optomechanical system to demonstrate off-resonant force and displacement sensitivity reaching 1.5dB below the SQL.  This achieves an outstanding goal in mechanical quantum sensing, and further enhances the prospects of using such devices for state-of-the-art force sensing applications.}
\end{abstract}
\newpage
The SQL can be derived through a straightforward analysis of the quantum noise sources in an ideal interferometric displacement measurement\cite{Clerk2010,Aspelmeyer2014,Braginsky1992,Braginskii1968}.  The premise of such a measurement is that one couples the position of an object (say, a harmonically bound mirror) to the phase of a coherent light field.  The uncertainty in this phase will scale inversely with the strength of the coherent field (as per the Heisenberg uncertainty relation).
This imprecision noise (`shot noise') constitutes an effective displacement noise $\ximp$, whose spectral density can be written as\cite{Aspelmeyer2014,Bowen2016}
\begin{equation}
\SxxImp = \frac{\xzpf^2}{4\Gmeas},
\end{equation}
where $\Gmeas$ is a measurement rate\cite{Bowen2016} characterizing the strength of the interaction, and $\xzpf=\sqrt{\hbar/2m\Om}$ is the root-mean-square amplitude of the oscillator's zero-point fluctuations (where $m$ is the mass, $\Om$ the resonance frequency, and $\hbar$ the reduced Planck's constant).  

\begin{figure}[tt!]
\begin{center}
\includegraphics[width=0.6\columnwidth]{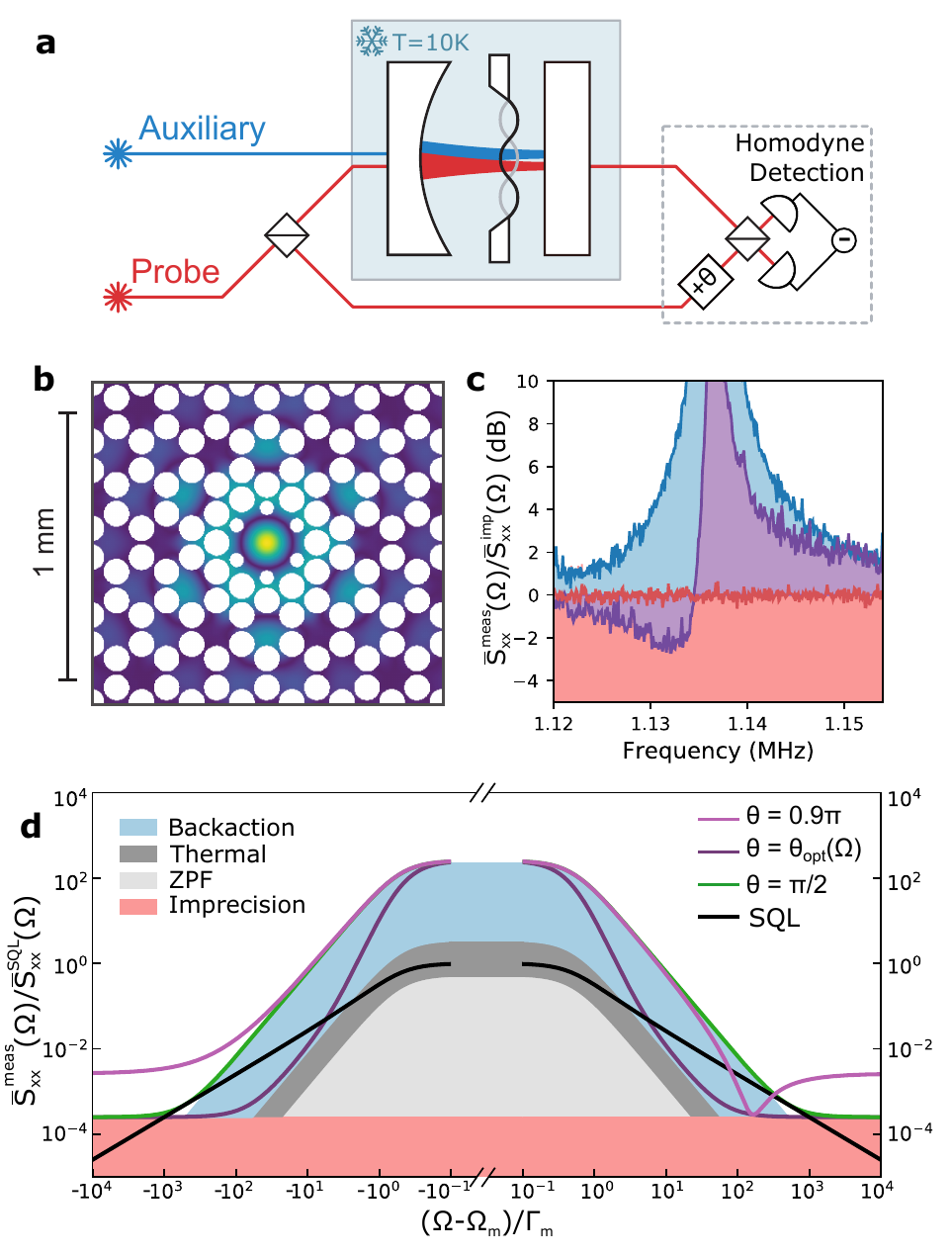}
\caption{{\bf Measuring beyond the standard quantum limit.}
{\bf a} Sketch of the experimental setup.
{\bf b} Simulated displacement pattern of the mechanical mode of interest.
{\bf c} Displacement spectra, normalized to the shot noise ($\SxxMeas(\Omega)/\SxxImp(\Omega)$), demonstrating ponderomotive squeezing.  Blue corresponds to a standard phase measurement ($\theta=\pi/2$), red indicates a shot noise measurement, and purple indicates a measurement of a rotated quadrature ($\theta\approx0.16\pi$), for which strong correlations are visible.
{\bf d} Cartoon schematic of an interferometric displacement measurement.  Shaded regions indicate different noise contributions, while the solid curves indicate the total spectra, under different measurement conditions: ``standard'' phase measurement (green), fixed ``non-standard'' quadrature measurement (light purple), frequency-dependent (``variational'') quadrature measurement (dark purple).  The black line indicates the SQL.}
\label{f:intro}
\end{center}
\end{figure}

In addition to this imprecision noise, there will also be `backaction' to the measurement, in the form of radiation pressure fluctuations ($\Fqba$). These fluctuations will be proportional to the measurement strength, with a spectral density given by
\begin{equation}
\SFFqba = \hbar^2\xzpf^{-2}\Gmeas.
\end{equation} 
Through the mechanical susceptibility $\chim(\Og) = m^{-1}\left(\Om^2-\Og^2-\imath\Gm\Og\right)^{-1}$ (where $\Gm$ is the mechanical energy damping rate), this produces displacement fluctuations, such that the total displacement noise added by the measurement process is given by:
\begin{equation}\label{eqn:sadd_nocorr}
\SxxAdd(\Omega) = \SxxImp + |\chim(\Og)|^2\SFFqba(\Omega).
\end{equation}
Recalling how these terms scale with measurement strength, we see the fundamental tradeoff between imprecision and backaction in a displacement measurement.  Minimizing with respect to the measurement strength, one finds that the minimum added noise occurs for $\Gmeas^{\mathrm{opt}}(\Og)=\xzpf^2(2\hbar|\chim(\Omega)|)^{-1}$, at which point the two contributions are equal, and the minimum added noise, known as the SQL, is given by 
\begin{equation}
\SxxSQL(\Og) \equiv \mathrm{min}\,\SxxAdd(\Omega) = \hbar |\chim(\Og)|
\end{equation}
The added imprecision and backaction noise will appear in addition to the oscillator's intrinsic motion, $\SxxInt$, (consisting of zero-point motion and thermal motion), such that the \textit{total} measured displacement noise is given by:
\begin{equation}
\SxxMeas(\Omega) = \SxxInt(\Omega) + \SxxAdd(\Omega)
\end{equation}
These various contributions to the measured spectrum are illustrated in Fig.~\ref{f:intro}.
Note that non-negligible thermal motion prevents measurements at the SQL on the mechanical resonance.  As such, much of the progress in approaching the SQL in optomechanical systems has relied on first reducing the thermal motion, via cryogenics and laser cooling\cite{LaHaye2004,Kampel2017}.  Alternatively, if one can achieve quantum backaction noise which dominates over the thermal noise (as in Fig.~\ref{f:intro}), then there exist frequencies away from resonance where it is still possible reach the SQL.  Indeed, the closest absolute approach to the SQL has recently been demonstrated in precisely this fashion\cite{Rossi2018}.  Such broadband, off-resonance displacement measurements are relevant for many practical sensing applications, including gravitational wave detection. 

While the SQL, as formulated above, does present a lower bound on the sensitivity of conventional interferometry, it does not correspond to a fundamental quantum limit.  Indeed, there are various ways to modify the premises of the measurement posed above, to enable sensitivities below the SQL\cite{Kimble2002,Corbitt2004}.  One well-known class of measurements avoids the required addition of backaction by measuring an observable which commutes with the system's Hamiltonian, making a `quantum nondemolition' (QND) or `backaction evading' (BAE) measurement\cite{Braginsky1980}.  Devices which couple either to the speed of a free mass\cite{Braginsky1990,Purdue2002} (instead of displacement), or to a single mechanical quadrature\cite{Thorne1978,Braginsky1978} satisfy this criteria, and indeed the latter has been demonstrated in several optomechanical systems\cite{Suh2014,Wollman2015b,Lecocq2015}, though with excess imprecision noise which prevented measurement below the SQL.  
Backaction-evading measurements of composite quadratures in multimode systems have also been explored in optomechanical\cite{Ockeloen-Korppi2016} and hybrid spin-mechanical\cite{Moller2017} systems.
Still other techniques rely on modification of the mechanical susceptibility\cite{Arcizet2006} (e.g. by creating an `optical spring') to surpass the bare-resonator SQL (albeit establishing a new SQL at the shifted frequency).

Yet another class of measurements relies instead on the exploitation of quantum correlations to deviate from the SQL.  Here, the underlying idea is that the derivation of the SQL assumed that the backaction and imprecision noises were uncorrelated (i.e. $\bar{S}_{x F}(\Og)=0$).  Without this assumption, equation~\eqref{eqn:sadd_nocorr} reads (See Methods):
\begin{equation}
    \SxxAdd(\Omega) = \SxxImp + |\chim(\Og)|^2\SFFqba(\Omega) + 2\mathrm{Re}\left[\chim(\Og)^*\bar{S}_{x F}(\Og)\right].
\end{equation}

By utilizing correlations in the measurement spectrum, it is possible to arrange for destructive interference of the quantum imprecision and backaction noise.  
This can be achieved either by  modifying the correlations of the input light (e.g. probing with squeezed instead of coherent light\cite{Unruh1982,Caves1981,Hoff2013,LIGO2011}), or by taking advantage of optomechanically-induced correlations in the output light\cite{Vyatchanin1995}.  
These correlations exist whenever backaction-dominated motion (arising from amplitude fluctuations of the optical field) is imprinted as phase fluctuations of that same field.  These correlated fluctuations are the same which enable ponderomotive squeezing\cite{Braginskii1967}.  To see these correlations, one simply needs to measure some mixed quadrature (containing both amplitude and phase fluctuations) of the optical field, as opposed to just making a phase measurement.
Of course, measuring quadratures other than the optical phase will sacrifice some of the mechanical signal (i.e. increase the imprecision), but for some optical quadratures, the reduction in noise outweighs the loss in signal in a certain frequency band.  This is illustrated in Fig.~\ref{f:intro}.  Compared to the normal phase measurement (corresponding to a quadrature angle $\theta=\pi/2$), a detection angle of $\theta=0.9\pi$ results in destructive interference of the backaction and imprecision near some particular frequency.  A complete implementation of this `variational' measurement actually envisions filtering the signal in such a way as to measure a frequency-dependent optical quadrature ($\theta_{\mathrm{opt}}(\Omega)$), such that this interference occurs over a broad frequency range above and below resonance.  This displacement sensitivity translates directly (via the susceptibility $\chi_m$) into force sensitivity, as we will see later.

We note that exploiting quantum correlations in this way requires efficient optical detection, in addition to strong backaction. Any optical losses spoil the needed interference by substituting uncorrelated vacuum noise.
Indeed, a recent work by Kampel et al. \cite{Kampel2017} has provided an extensive theoretical and experimental study of variational displacement measurements, but excess imprecision noise limited the sensitivity of the measurement to 0.9~dB above the SQL.

In this sense, while the principles underlying both BAE and variational techniques have been demonstrated, experiments to-date have fallen short of proving the original intent of these techniques, and interferometric force and displacement measurements beyond the SQL has remained an open experimental challenge\cite{Giovannetti2004}.  

Here, we are able to address this challenge via an optomechanical system which permits strong, efficient quantum measurements, thanks to ultracoherent mechanical devices and carefully optimized optical readout. 
The mechanical system is based on a 3.6mm$\times$3.6mm$\times$20nm $\mathrm{Si_3N_4}$ membrane, patterned with a honeycomb lattice of holes to create a phononic crystal (PnC) with an acoustic bandgap near 1~MHz. 
At the center of the membrane, a defect supports several localized, out-of-plane vibrational modes, whose frequencies lie within the bandgap.  The PnC shields these modes from radiative loss, while simultaneously providing a ``soft clamp'', which dramatically reduces the mechanical loss\cite{Tsaturyan2017}. 
In this work, we focus on a single mode with resonance frequency $\Omega_m/2\pi=$~1.135~MHz and quality factor $Q=\Om/\Gm=1.03\times10^9$ at temperature $T=$~10~K. The motion of the membrane is dispersively coupled to the resonance frequency of a Fabry-P\'erot cavity mode (linewidth $\kappa/2\pi=$~16.2~MHz) at a characteristic vacuum coupling rate $g_0/2\pi=$~120.7~Hz. 
By driving the cavity mode near resonance, we populate a coherent field with average occupation $\bncav$, which leads to a field-enhanced, linearized coupling at rate $g=g_0 \sqrt{\bncav}$. The cavity is highly over-coupled in transmission, such that 95\% of the intracavity photons emerge towards the detector. 
We monitor the optical quadrature fluctuations of this output beam via a balanced homodyne detector, which can measure at an arbitrary quadrature angle, $\theta$.
By optimizing the output optics and homodyne detection, we achieve a total detection efficiency of $\etad=$~77\%.
In addition to a resonant probe beam, we also use an auxiliary laser, addressing a separate cavity mode.  This beam is used simultaneously for sideband cooling and as an actuator beam for feedback cooling of low-frequency out-of-bandgap membrane modes (for stabilization purposes\cite{Rossi2018}). 
This auxiliary beam also exerts a small amount of quantum backaction on the membrane, which we account for as an effective thermal bath (since it represents a source of uncorrelated force noise from the perspective of the probe beam).
In the results which follow, it will be important that we know the transduction factor between mechanical displacement and photocurrent, for calibrating spectra.  This is accomplished by a well-established phase modulation technique\cite{Gorodetsky2010}.  This technique relies on knowledge of the optomechanical coupling rate, $g_0$, which we obtain through two independent, cross-checked calibration methods (see Methods).  We note that through these methods, we arrive at a calibration factor which enables robust comparison to the SQL, depending only on a few well-known system parameters (see Methods).
\begin{figure}[ht]
\begin{center}
\includegraphics[scale=1]{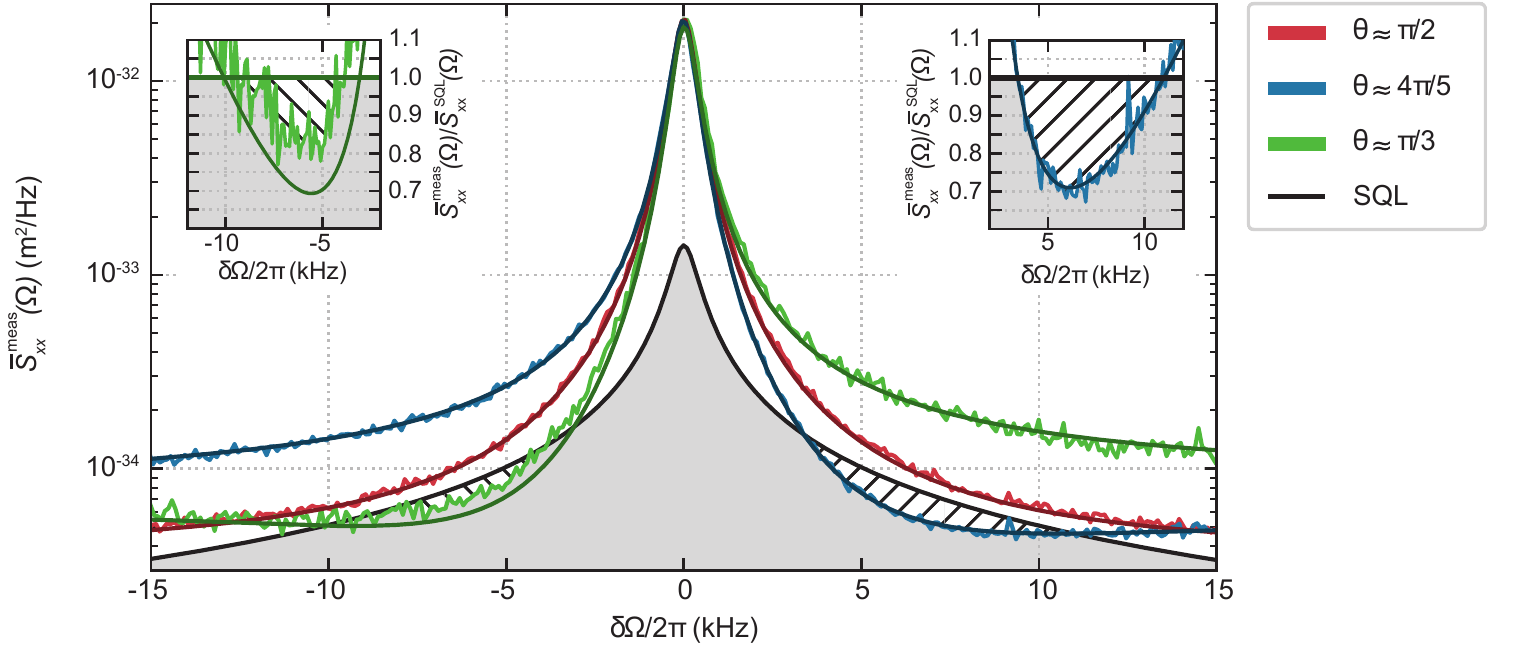}
\caption{{\bf Measuring displacement below the SQL.} Calibrated displacement spectra as measured at different quadrature angles of the homodyne receiver.  The red curve corresponds to a `conventional' ($\theta\approx\pi/2$) measurement, while blue and green correspond to quadratures which offer enhanced sensitivity above and below resonance, respectively.  Solid dark color lines are fits to a complete model (see Method). The solid black line is the SQL of the effective oscillator. In the insets, a zoom of the hatched regions below and above resonance, where the noise level decreases below the SQL. Abscissa refers to frequency relative to the effective mechanical resonance $\delta\Omega = \Og-\Oeff$.
\label{f:subSQLspectra}}
\end{center}
\end{figure}

As noted previously, to exploit broadband correlations in the probe field, it is necessary that quantum backaction is the dominant source of force noise driving the resonator (as opposed to the thermal bath with mean phonon occupation $\bnth$).  Quantitatively, this corresponds to reaching a ``quantum cooperativity'' $\Cq=\Gqba/\gamma\gtrsim 1$, where $\gamma$ is the thermal decoherence rate of the oscillator and $\Gqba=4g^2/\kappa$ is the quantum backaction decoherence rate.  Moreover, it is important that the total detection efficiency is close to unity.  A simple ponderomotive squeezing experiment (Fig.~\ref{f:intro}c) demonstrates that both of these are indeed satisfied in this system. We probe the motion with $\Cq=17.3$, and detect a homodyne quadrature corresponding to $\theta=0.16~\pi$. The strong correlations produce the observed asymmetric Fano resonance.  

Now, we examine what effect these correlations have on our displacement sensitivity, and compare with the SQL.  Here, the SQL is defined as $\SxxSQL(\Og) = \hbar |\chieff(\Og)|$, where $\chieff(\Og)$ is the effective mechanical susceptibility resulting from dynamical backaction of the (slightly red-detuned) probe and auxiliary lasers\cite{Clerk2015} (See Methods).  Figure~\ref{f:subSQLspectra} shows three displacement spectra (for $\Cq=17.3$), one measured in a ``standard'' ($\theta\approx\pi/2$) configuration, and two measured at homodyne angles which are optimized for displacement sensitivity above and below resonance.  We see that the correlations enable sub-SQL sensitivity in both regions, with the best sensitivity occurring at $\Og-\Om^\mathrm{eff} = 2\pi\times5.9$~kHz.  Here, the total noise, including even thermal and zero-point fluctuations, is 1.5~dB below the SQL.  The solid lines indicate predictions of standard optomechanical input-output theory\cite{Aspelmeyer2014}, in which three parameters ($g$, $\theta$, and cavity detuning $\Delta$) are adjusted to fit the data  (see Methods).  This is to account for small experimental drifts in intracavity power and laser detuning while we systematically vary the detection angle, $\theta$.

\begin{figure}
\begin{center}
\includegraphics[width=0.95\columnwidth]{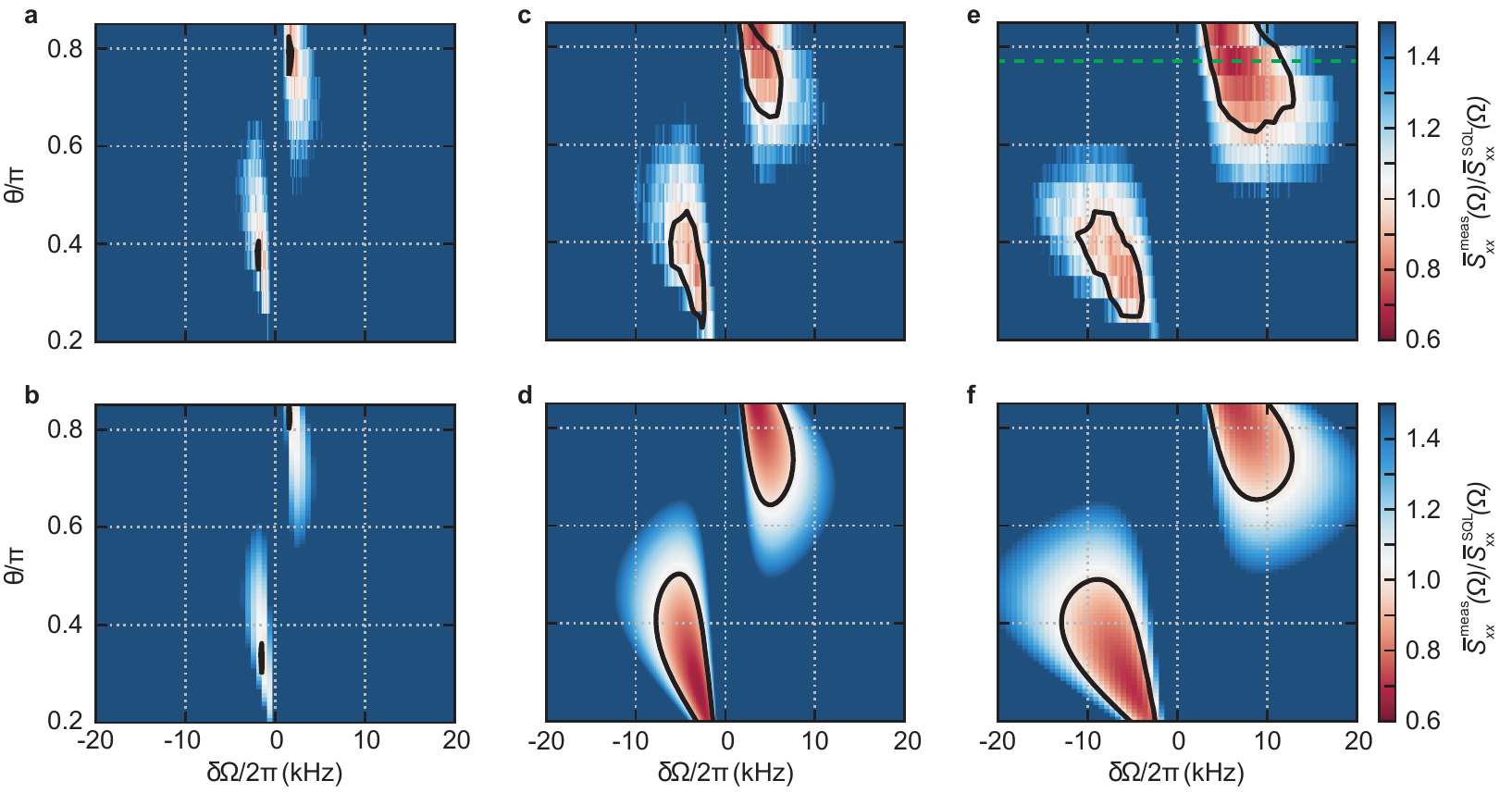}
\caption{{\bf Opening broadband regions of sub-SQL sensitivity.}
{\bf a, c, e} Calibrated displacement spectra (for $\Cq=$~\{4.6, 8.8, 17.3\}, respectively) near the mechanical resonance frequency $\Oeff$, for varying measurement quadrature ($\theta$). The spectra are divided by the SQL, such that red regions indicate sub-SQL sensitivity. Black contour lines corresponds to $\bar S_{xx}^\mathrm{meas}=\SxxSQL$. The dashed green line in {\bf e} corresponds to the blue spectrum in Fig.~\ref{f:subSQLspectra}.
{\bf b, d, f} Independent predictions from the theoretical model (see Methods).
\label{f:systematic}}
\end{center}
\end{figure}

Recalling Fig.~\ref{f:intro}, we note that different measurement angles optimize the sensitivity at different frequencies, and that larger backaction results in a larger frequency region in which correlations can be exploited.  We demonstrate these dependencies in Fig.~\ref{f:systematic}, which shows spectra (normalized to the SQL) measured at various angles ($\theta\in~[0.1\pi, 0.9\pi]$) and for $\Cq=$\{4.6, 8.8, 17.3\}.  We see that the bandwidth of sub-SQL performance is improved at larger values of $\Cq$.  
Figure~\ref{f:systematic} shows theoretical spectra based purely on independent parameters, and we note broad consistency between the data and model throughout the parameter space.  We note that the measurements generally show degraded performance on the lower-frequency side of resonance, which we attribute to the presence of extraneous cavity noise.

\begin{figure}
\begin{center}
\includegraphics[scale=1]{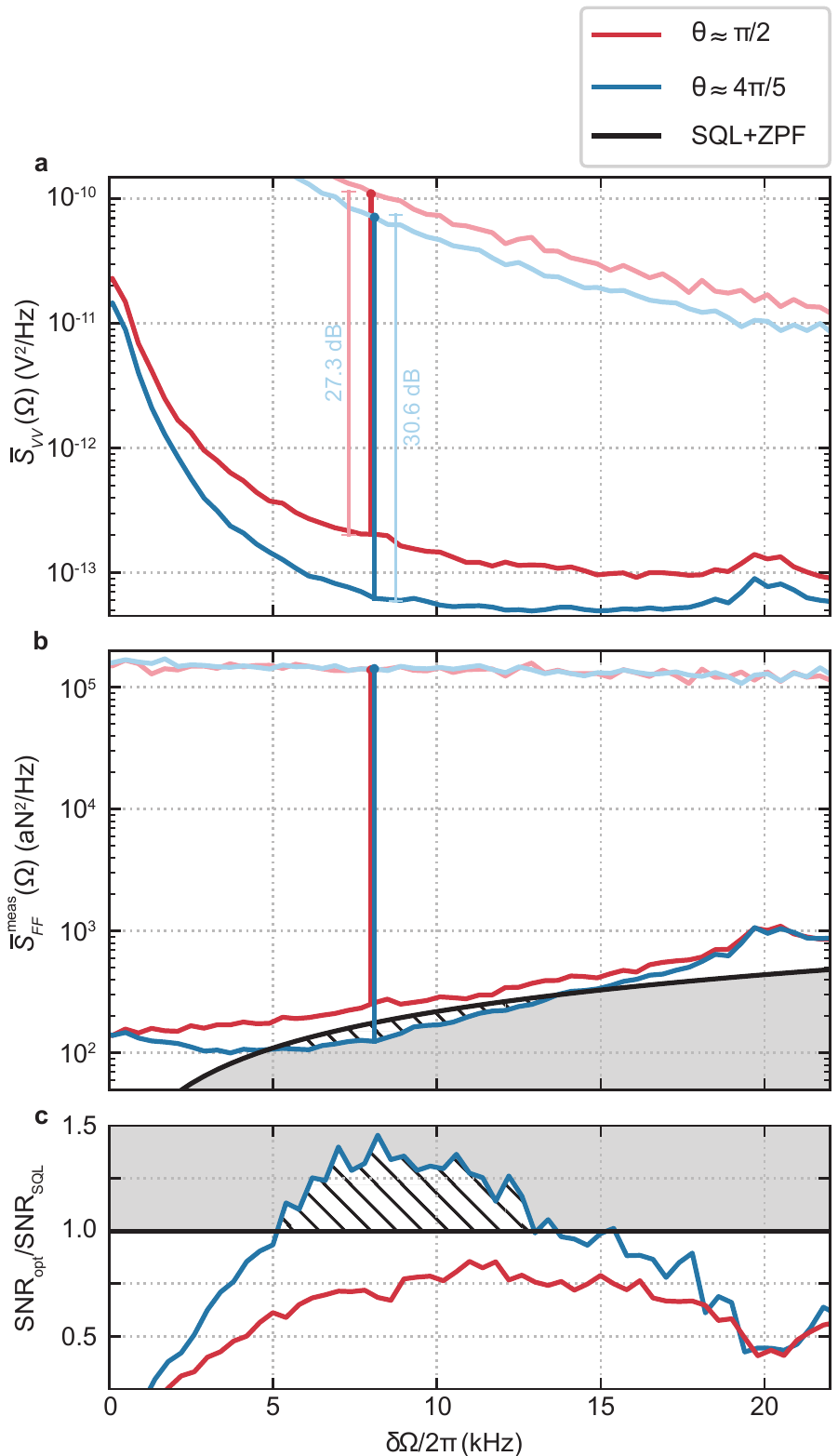}
\caption{{\bf Quantum-enhanced force sensing beyond the SQL.}
{\bf a,} Raw photocurrent spectra (dark lines) and driven response (light lines), for $\theta\approx\pi/2$ (red) and $\theta\approx4\pi/5$ (blue).  The response at $\delta\Og/2\pi=$8.2kHz is highlighted for emphasis
{\bf b,} Same noise spectra and driven response measurements from {\bf a}, calibrated in units of force noise.
{\bf c,} Signal-to-noise ratio (SNR), relative to the SNR of a measurement in which $\SxxAdd=\SxxSQL$.
\label{f:force_snr}}
\end{center}
\end{figure}

Sensitive displacement measurements are the foundation of sensitive force measurements.  Here, we demonstrate proof-of-principle force measurements to illustrate how the signal-to-noise (SNR) ratio in such a measurement is improved through variational techniques, beyond what is possible at the SQL.  We particularly emphasize how the signal and noise respond differently to measurement quadrature changes.  By modulating the amplitude of the auxiliary laser, we apply a coherent, classical radiation pressure force on the mechanical resonator.  We measure the response via homodyne detection as before, and compare with noise measurements as above.  Figure~\ref{f:force_snr} shows signal and noise measurements, in both raw voltage units and after using the calibration tone (and $\chieff$) to calculate force spectra.  The driven response at all frequencies is shown, with the signal at one frequency ($\Omega_0$) emphasized for comparison.  From the raw spectra (Fig.~\ref{f:force_snr}a), we note that between $\theta\approx\pi/2$ and $\theta\approx4\pi/5$, the signal at $\Omega_0-\Oeff=2\pi\times8.2$~kHz decreases by 1.9~dB (due to reduced motional transduction), while the noise level drops by 5.2~dB.  The reduced transduction is accounted for with our calibration method, such that in Fig.~\ref{f:force_snr}b we find equal force signals from both configurations (and for all frequencies), but still with reduced noise levels.  The signal-to-noise ratio (SNR) is summarized in Fig.~\ref{f:force_snr}c, normalized to the SNR possible for a measurement operating at the SQL.  (Note that we include in this noise floor the zero-point-fluctuations (ZPF), as they represent another fundamental noise source.  In practice, this ZPF term\cite{Clerk2004} has a negligible impact in the frequency region where we find sub-SQL sensitivity). As in the noise-only measurements above, we find that the variational measurement enables SNR values exceeding what the SQL would allow over a 8~kHz bandwidth.  Finally, we note that with the application of sub-SQL techniques, our force sensitivity, at $\Omega_0$, reaches $(11.2\,\mathrm{aN/\sqrt{Hz}})^2$ -- a number which can be improved significantly in low-mass versions of this system, optimized for force sensing.

The results presented here demonstrate that the position, even of a macroscopic object, can now be probed with a sensitivity reaching the limit which quantum mechanics imposes on conventional measurement.  Moreover, we see how one can go beyond these limits by better using the quantum resources of the measurement (in this case, strong correlations).  Indeed, in the sense that ponderomotive squeezing reflects entanglement between probe light sidebands\cite{Zippilli2015}, the technique demonstrated here can be considered a form of entanglement-enhanced measurement.  The strong quantum interactions possible in this system should also allow explorations of similar quantum-enhanced measurement techniques, including synodyne detection\cite{Buchmann2016} and single-quadrature measurements\cite{Clerk2008}.  It has also been shown that the limits of measurement-based quantum feedback (as was recently demonstrated in this system\cite{Rossi2018}) can be improved by exploiting variational techniques like the one shown here\cite{Habibi2016}.  Finally, the enhanced force sensitivity could be directly applicable in state-of-the-art force sensing applications\cite{Poggio2010}, opening new regimes of precision measurement.
%
%
%
%
%
%
%
%

\clearpage
\newpage
\section*{{\Large Method}}
\section{Model}
\begin{figure}[h!]
\begin{center}
\includegraphics[scale=1]{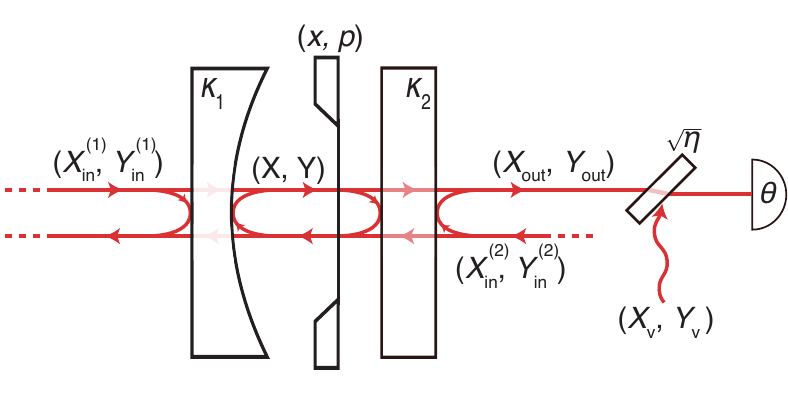}
\caption{{\bf Optomechanical cavity.} A coherent field drives a cavity mode from the left port. An homodyne receiver is set up to measure a quadrature $\theta$ of the transmitted field, through the right port of the cavity. Optical losses are modelled with uncorrelated vacuum entering from a beam-splitter with transmissivity $\eta$.
\label{f:theory}}
\end{center}
\end{figure}
The homodyne spectra from the main text can be modelled within the standard framework of linearized optomechanics\cite{Bowen2016,Aspelmeyer2014,Kampel2017}.  In this way, the system dynamics are described by the following set of linear, coupled differential equations:
\begin{subequations}\label{e:qle_linear}
\begin{align}
    \dot{\hat{X}} &= -\frac{\kappa}{2}\hat{X}-\Delta\hat{Y}+\sqrt{\kappa_1}\hat{X}^{(1)}_\mathrm{in}+\sqrt{\kappa_2}\hat{X}^{(2)}_\mathrm{in},\\
    \dot{\hat{Y}} &= -\frac{\kappa}{2}\hat{Y}+\Delta\hat{X}+2 g\frac{1}{\sqrt{2}\xzpf} \hat{x}+\sqrt{\kappa_1}\hat{Y}^{(1)}_\mathrm{in}+\sqrt{\kappa_2}\hat{Y}^{(2)}_\mathrm{in},\\
    m\ddot{\hat{x}} &= -m\Om^2\hat{x} -m\Gm\dot{\hat{x}} + 2g\frac{\hbar}{\sqrt{2}\xzpf}\hat{X} + \sqrt{\Gm}\Fth,
\end{align}
\end{subequations}
where $\hat{X}$, $\hat{Y}$ and $\hat{x}$ describe, respectively, the cavity amplitude and phase fluctuations and the mechanical position (Fig.~\ref{f:theory}).  The loss rates through the two mirrors are given by $\kappa_1$ and $\kappa_2$, and the total cavity decay rate is $\kappa=\kappa_1+\kappa_2$.
The thermal Langevin force ($\Fth$) satisfies $\langle\Fth(t)\Fth(t')\rangle=2 m\hbar \Om(\bnth+1/2)\,\delta(t-t')$.
The terms $\hat{X}_\mathrm{in}^{(i)}$ and $\hat{Y}_\mathrm{in}^{(i)}$ represent, respectively, vacuum amplitude and phase fluctuations entering the $i$-th port of the cavity. Their correlations are
\begin{subequations}
\begin{align}
    \langle\hat{X}_\mathrm{in}^{(i)}(t)\hat{X}_\mathrm{in}^{(j)}(t')\rangle&=\langle\hat{Y}_\mathrm{in}^{(i)}(t)\hat{Y}_\mathrm{in}^{(j)}(t')\rangle=\frac{1}{2}\,\delta_{ij}\,\delta(t-t'),\\
    \langle\hat{X}_\mathrm{in}^{(i)}(t)\hat{Y}_\mathrm{in}^{(j)}(t')\rangle&=\frac{\imath}{2}\,\delta_{ij}\,\delta(t-t').
\end{align}
\end{subequations}
This system has been considered in many excellent resources\cite{Bowen2016,Kampel2017}; for our starting point, we will consider a general quadrature of the transmitted output optical field,
\begin{equation}\label{eqn:general_quad}
\hat{X}_\mathrm{out}^\theta(\Og) = \hat{X}_\mathrm{in}^\theta + f^\theta(\Og) \hat{x}(\Og),
\end{equation}
where the mechanical position $\hat{x}$ is transduced into quadrature fluctuations via the quadrature-dependent function $f^\theta(\Og)$; $\hat{X}^\theta_\mathrm{in}$ is a combination of input vacuum quadratures whose spectrum\footnote{We use a double-sided spectral convention, where $S_{AB}(\Og) = \int_{-\infty}^{\infty} dt e^{\imath \Og t} \langle \hat{A}^\dagger(t) \hat{B}(0)\rangle$. Symmetrisation, i.e. $\bar{S}_{AB}(\Og) = (S_{AB}(\Og) + S_{AB}(-\Og))/2$ is indicated by a overbar.}
is $S_{XX}^{\theta,\mathrm{in}}=1/2$. We note here that the angle $\theta$ is referred to the intracavity field, assumed real as phase reference.
The transduction function is given by,
\begin{equation}
f^\theta(\Og) = -\imath \frac{g}{\sqrt{2}\xzpf} \sqrt{\kappa_2} \left[\chic(\Og) e^{\imath\theta}-\chic(-\Og)^* e^{-\imath\theta}\right],
\end{equation}
where $\chic(\Og) = \left[\kappa/2-\imath\left(\Delta + \Og\right)\right]^{-1}$ is the cavity susceptibility.
The position of the mechanical oscillator $\hat{x}$ is
\begin{equation}
\hat{x}(\Og) = \chieff(\Og)\left[\Fth(\Og) + \Fqba(\Og)\right]
\end{equation}
and it is driven by two forces: a Langevin force ($\Fth$) from the thermal bath and a quantum backaction force ($\Fqba$) from radiation pressure shot noise. They induce motion via the susceptibility $\chieff(\Og)^{-1} = \chim(\Og)^{-1} -\imath 2 g^2 m\Om \left[\chic(\Og) - \chic(-\Og)^*\right]$, effectively modified by dynamical backaction\cite{Bowen2016, Aspelmeyer2014}.

The output quadrature in equation~\eqref{eqn:general_quad} is measured by a balanced homodyne receiver. If optical losses are present, the unitless photocurrent becomes $I = \sqrt{\etad}\hat{X}_\mathrm{out}^\theta + \sqrt{1-\etad}\hat{X}_\mathrm{v}$, with $\hat{X}_\mathrm{v}$ uncorrelated vacuum noise and $\etad$ detection efficiency. To express the photocurrent in units of position, we divide $I$ by $\sqrt{\etad}f^\theta(\Og)$. Thus, the measured displacement is $\hat{x}_\mathrm{meas} = \hat{x}_\mathrm{imp}+\hat{x}$, where $x_\mathrm{meas} = I/\sqrt{\etad}f^\theta(\Og)$  and $x_\mathrm{imp} = (\sqrt{\etad}X^\theta_\mathrm{in}+\sqrt{1-\etad}\hat{X}_\mathrm{v})/\sqrt{\etad}f^\theta(\Og)$. The symmetrized spectrum is 
 \begin{equation}\label{eqn:spectrum_total}
     \bar{S}_{xx}^\mathrm{meas}(\Og) = \bar{S}_{xx}^\mathrm{imp}(\Og) + |\chieff(\Og)|^2\left[\SFFth(\Og) + \SFFqba(\Og)\right] + 2 \mathrm{Re}\left[\chieff(\Og)^* \bar{S}_{x F}(\Og) \right].
 \end{equation}
Different contributions can be identified. The input optical quadrature produces an apparent background (imprecision noise) given by
\begin{equation}\label{eqn:sxximp}
     \bar{S}_{xx}^\mathrm{imp}(\Og) = \frac{\xzpf^2}{g^2 \etad\kappa}\frac{1}{|\chic(\Og)|^2 + |\chic(-\Og)|^2 - 2\mathrm{Re}\left[e^{\imath2\theta}\chic(\Og)\chic(-\Og)\right]}.
\end{equation}
The quantum backaction force spectrum is given by
\begin{equation}
	\SFFqba(\Og) = \frac{\hbar^2}{2\xzpf^2} g^2\kappa\left(|\chi_c(\Omega)|^2+|\chi_c(-\Omega)|^2\right),
\end{equation}
while the thermal Langevin force is $\SFFth(\Og) = m \Gm \hbar\Om  (2 \bnth+1) $. Finally, the last term appears because, generally speaking, the quantum backaction force and the imprecision noise are caused by partially the same fluctuations. Properly taking these correlations into account gives
\begin{equation}
     \bar{S}_{x F}(\Og) = \frac{\hbar}{2\imath}\frac{\chi_c(\Omega)e^{\imath\theta}+\chi_c(-\Omega)^*e^{-\imath\theta}}{\chic(\Og)e^{\imath\theta}-\chic(-\Og)^*e^{-\imath\theta}}.
\end{equation}
Note that by setting this correlation term to zero in equation~\ref{eqn:spectrum_total}, and optimizing the combined imprecision and backaction, one can find the SQL as used in the main text: $\SxxSQL(\Og) = \hbar |\chi_{\mathrm{eff}}(\Omega)|$.

In the resonant case ($\Delta=0$) and for a ``bad cavity'' ($\kappa\gg\Om$) the previous contributions can be simplified. The imprecision noise becomes
\begin{equation}
     \bar{S}_{xx}^\mathrm{imp}(\Og) \approx \frac{\xzpf^2}{16 g^2 \etad/\kappa}\frac{1}{\sin^2{\theta}},
\end{equation}
which is minized for a phase measurement, i.e. $\theta=\pi/2$. The quantum backaction force, indipendent from the measured quadrature, becomes
\begin{equation}
	\SFFqba(\Og) \approx \frac{\hbar^2}{2\xzpf^2} \frac{8 g^2}{\kappa}
\end{equation}
and the correlation term is
\begin{equation}
     \bar{S}_{x F}(\Og) \approx -\frac{\hbar}{2}\cot{\theta}.
\end{equation}

 \section{Model fit}
\begin{figure}
\begin{center}
\includegraphics[scale=1]{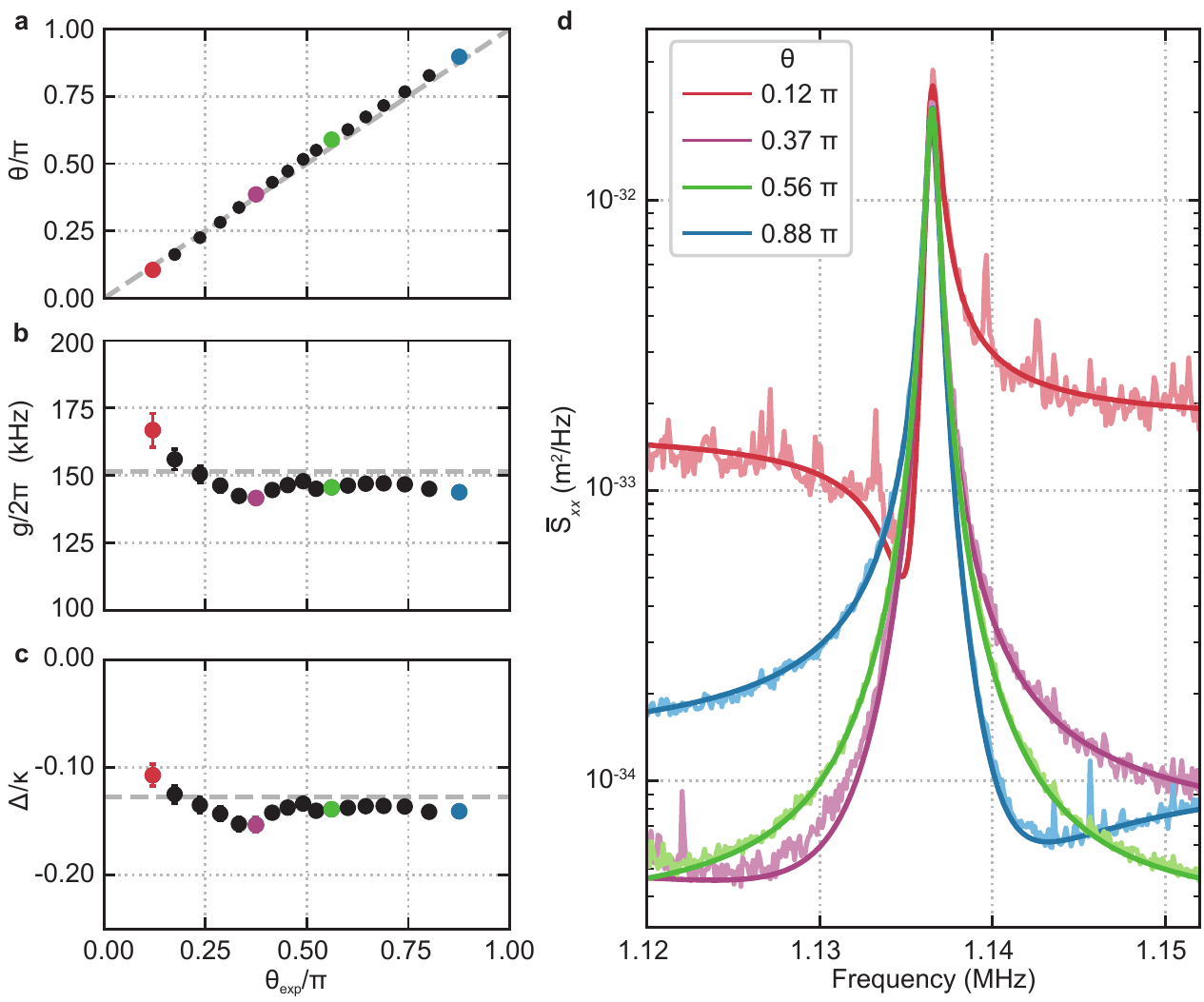}
\caption{{\bf Fit of displacement spectra.} 
{\bf a--c}, Fitted parameters from the spectra of the measured photocurrent. Dashed gray lines are predictions from independent measurements.
{\bf d}, Example of measured spectra (light) and fit (dark). The corresponding fitted parameters are marked, with the same colours, in {\bf a--c}.
\label{f:fit_example}}
\end{center}
\end{figure}

The model in equation~\eqref{eqn:spectrum_total} is used to fit measured photocurrent, calibrated in absolute displacement units (see below). To account for slow drifts of parameters during the experiment, we fit the homodyne angle $\theta$, the coupling $g$, and the detuning $\Delta$ (Fig.~\ref{f:fit_example}a-d). The asymmetric lineshape due to correlations is reproduced by the fit model, as shown in Figure~\ref{f:fit_example}d. All the other parameters which enter in equation~\eqref{eqn:spectrum_total} are fixed by independent measurement and predictions.

\begin{figure}
\begin{center}
\includegraphics[scale=1]{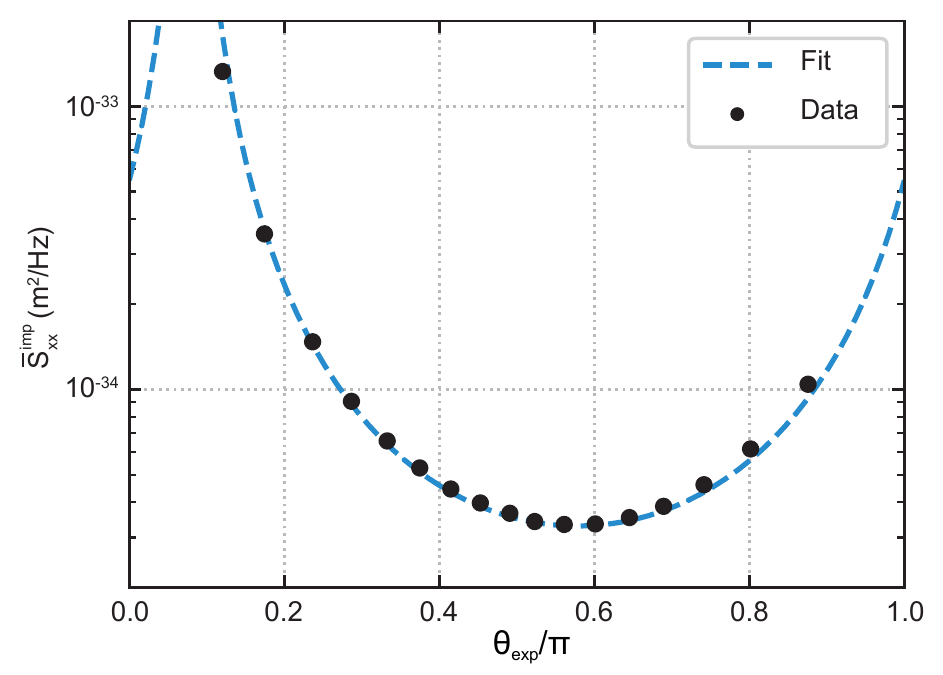}
\caption{{\bf Calibrated imprecision noise.} By means of phase modulation technique, the homodyne receiver shot noise level is calibrated into displacement imprecision noise, for different measured quadratures $\theta$. From a fit the detection efficiency $\eta$ is derived.
\label{f:homodyne_imp}}
\end{center}
\end{figure}

In particular, the detection efficiency $\etad$ is derived from fitting the calibrated shot noise level ($\SxxImp$) as a function of $\theta$ (equation~\eqref{eqn:sxximp}), as shown in Fig.~\ref{f:homodyne_imp}.

We notice a strong correlation between the parameters $\tilde{g}$ and $\tilde{\Delta}$ from the fit. This can be explained with variation of the input probe power. In fact, it directly changes the intracavity photon number $\bncav$ and, thus, $g$. At the same time, a power drift changes the setpoint at which the detuning $\Delta$ is stabilized through a Pound-Drever-Hall lock with a non-zero offset.

\section{Calibration}
When making absolute comparisons to the SQL, careful calibration of the noise spectra is of key importance.  Here we detail our approach, which is fundamentally based on using quantum backaction as a calibrated thermal bath.  Additional details about this method are also available in our previous work\cite{Rossi2018}. 

The calibration is composed of two steps. First, we conduct a sideband cooling experiment, in order to establish a well-known temperature reference.
This involves monitoring the motion with a weak probe beam, while an auxiliary beam (with detuning $\deltaaux=-0.33\kappa_\mathrm{aux}$, where $\kappa_\mathrm{aux}$ is the auxiliary cavity mode linewidth) is used to sideband cool the mechanical mode, reaching an asymptotic phonon occupancy ($\bnmin$) set by quantum backaction (QBA). This backaction provides a well-known bath to define the temperature of the mechanics. This motion translates to cavity frequency noise via the optomechanical coupling, $g_0$ as follows: 
\begin{equation}
    \langle\delta \Omega^2\rangle_\mathrm{mech}^\mathrm{QBA}=2 g_0^2 \left(\bnmin+\frac{1}{2}\right),
\end{equation}
The transduction factor of this frequency noise into voltage noise is measured via a calibrated phase modulation tone at frequency $\Omega_{\mathrm{cal}} \approx\Om $:
\begin{equation}
\label{e:K}
    K = \frac{\langle\delta V^2\rangle_\mathrm{ cal}^\mathrm{QBA}}{\langle\delta \phi^2\rangle_\mathrm{cal}^\mathrm{QBA}}=\frac{\langle\delta V^2\rangle_\mathrm{ cal}^\mathrm{QBA}}{\phi^2/2},
\end{equation}
where $\phi$ the phase modulation depth of the calibration tone. Thus, the measured voltage variance due to mechanical motion can be expressed in terms of frequency variance as
\begin{equation}
    \langle\delta V^2\rangle_\mathrm{mech}^\mathrm{QBA}=K\langle\delta\phi^2\rangle_\mathrm{mech}^\mathrm{QBA}=K\langle\delta\Omega^2\rangle_\mathrm{mech}^\mathrm{QBA}/\Omega_m^2=\frac{K2 g_0^2}{\Omega_m^2} \left(\bnmin+\frac{1}{2}\right).
\end{equation}
Substituting $K$ from equation~\eqref{e:K}, and solving for $g_0$, we have
\begin{equation}
\label{e:g0}
    g_0=\sqrt{\frac{\langle\delta V^2\rangle_\mathrm{mech}^\mathrm{QBA}}{\langle\delta V^2\rangle_\mathrm{ cal}^\mathrm{QBA}}\frac{\Omega_m^2 \phi^2/2}{2(\bnmin+1/2)}}.
\end{equation}

From this measurement, we find $g_\mathrm{0}/2\pi=$ 120.7~Hz. Actually we fit $g_0$ from the entire sideband cooling dataset (various cooling powers), to provide a more robust $g_0$ estimate, as well as an estimate of the thermal bath temperature. As a cross-check, we also conduct an optomechanically-induced transparency (OMIT) experiment, which also calibrates $g_0$, but relying on a different set of parameter assumptions.  The OMIT calibration only differs from the backaction calibration by 3\% ($g_0/2\pi=$124.7~Hz).

With $g_0$ obtained, we can now use the same calibration technique to convert arbitrary voltage spectra ($\bar{S}_{VV}$) into displacement spectra ($\bar{S}_{xx}^\mathrm{meas}$)\cite{Gorodetsky2010}[Gorodetksy2010]:
\begin{equation}
    \bar{S}_{xx}^\mathrm{meas}=\frac{\xzpf^2 \Om^2 \phi^2/2}{g_0^2 \langle\delta V^2\rangle_\mathrm{cal}^\mathrm{meas}}\bar{S}_{VV},
\end{equation}
where $\langle\delta V^2\rangle_\mathrm{cal}^\mathrm{meas}$ is the voltage variance of the calibration tone \emph{in that particular measurement}.  Note that this is different from $\langle\delta V^2\rangle_\mathrm{cal}^\mathrm{QBA}$, which refers specifically to the calibration tone voltage during the QBA calibration measurement.
\\
Using equation~(\ref{e:g0}) we recast this conversion factor as:
\begin{equation}
    \bar{S}_{xx}^\mathrm{meas}=\frac{\xzpf^2 (2\bnmin+1)}{\langle\delta V^2\rangle_\mathrm{mech}^\mathrm{QBA} }\frac{\langle\delta V^2\rangle_\mathrm{cal}^\mathrm{QBA}}{\langle\delta V^2\rangle_\mathrm{cal}^\mathrm{meas}} \bar{S}_{VV}.
\end{equation}
We note that even the modulation depth of the calibration tone has now cancelled out.  The only parameters which remain are several precisely-known frequencies, $\bnmin$ (which itself depends only on three well-known variables: $\kappaaux,\deltaaux$, and $\Om$\textbf{}), $\xzpf$, and a few voltages which are read directly from various spectra.  The only remaining uncertainty might be in $\xzpf$, due to its dependence on the oscillator mass, but since $\bar{S}_{xx}^\mathrm{SQL}=\hbar |\chi_\mathrm{m}|=\hbar m^{-1} /\sqrt{(\Om^2-\Og^2)^2+\Gm^2\Og^2}$ has the same mass-dependence, our figure-of-merit comparisons remain robust:
\begin{equation}
    \frac{\bar{S}_{xx}^\mathrm{meas}}{\bar{S}_{xx}^\mathrm{SQL}}=\frac{\Om (\bnmin+1/2) \sqrt{(\Om^2-\Og^2)^2+\Gm^2\Og^2}}{\langle\delta V^2\rangle_\mathrm{mech}^\mathrm{QBA} \Omega_\mathrm{cal}^2}\frac{\langle\delta V^2\rangle_\mathrm{cal}^\mathrm{QBA}}{\langle\delta V^2\rangle_\mathrm{cal}^\mathrm{meas}} \bar{S}_{VV}.
\end{equation}

%
%
\clearpage
\newpage
\bibliography{subSQLbibliography}

\end{document}